\documentstyle[prb,aps,epsf]{revtex}
\begin{document}
\title{
Physical properties of misfit-layered (Bi,Pb)-Sr-Co-O system: 
Effect of hole doping into triangular lattice formed by low-spin Co ions
} 

\author{
T. Yamamoto~\cite{A} and K. Uchinokura~\cite{B}
}
\address{
\it Department of Applied Physics, The University of Tokyo, 
7-3-1 Hongo, Bunkyo-ku, Tokyo 113-8656, Japan
}

\author{
I. Tsukada
}
\address{
\it Electrical Physics Department, Central Research Institute of 
Electric Power Industry, 2-11-1 Iwato-kita, Komae, Tokyo 201-8511, Japan
}

\date{\today}
\maketitle
\begin{abstract}
Pb-doping effect on physical properties of misfit-layered (Bi,Pb)-Sr-Co-O system, 
in which Co ions form a two-dimensional triangular lattice, was 
investigated in detail by electronic transport, magnetization and specific-heat 
measurements. Pb doping enhances the metallic behavior, suggesting that carriers are doped. 
Pb doping also enhances the magnetic correlation in this system and increases the magnetic transition temperature. 
We found the existence of the short-range magnetic correlation far above the transition temperature, 
which seems to induce the spin-glass state coexisting with the ferromagnetic long-range order at low temperatures. 
Specific-heat measurement suggests that the effective mass of the carrier in (Bi,Pb)-Sr-Co-O 
is not enhanced so much as reported in NaCo${}_2$O${}_4$. 
Based on these experimental results, we propose a two-bands model 
which consists of narrow $a_{1g}$ and rather broad $e'{}_g$ bands. 
The observed magnetic property and magnetotransport phenomena are explained well by this model. 
\end{abstract}

\section{Introduction} 
\begin{figure}
\epsfxsize=8.5cm
\centerline{\epsfbox{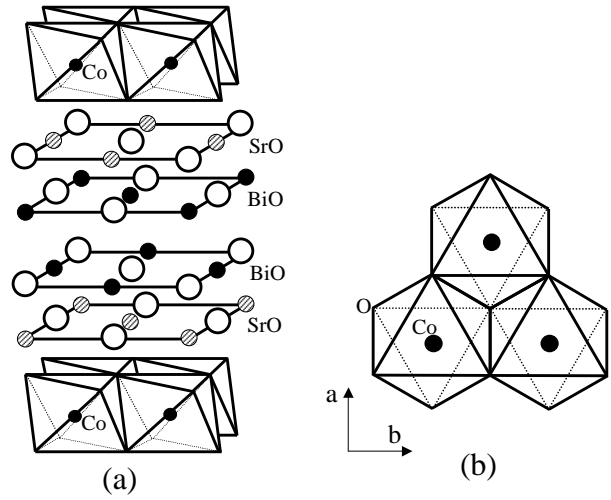}}
\caption{
(a) A schematic picture of the crystal structure of Bi-Sr-Co-O. 
(b) A projected view of hexagonal CoO${}_2$ layer into the $ab$ plane. 
} 
\label{cryst}
\end{figure}

The discovery of large thermoelectric power in NaCo${}_2$O${}_4$~(Ref.~\onlinecite{Terasaki97}) 
has made us conscious of the potential of 3$d$ transition-metal oxides 
as thermoelectric materials. Since then, many researchers have begun to 
search for materials that have still better thermoelectric properties. 
Recently, it is reported that (Bi,Pb)-Sr-Co-O 
and Ca${}_3$Co${}_4$O${}_9$ also show large thermoelectric 
power.~\cite{Woosuck00,Funahashi00,Itoh00,Masset00,Miyazaki00} 
The former is one of the variations of compounds of Bi-Sr-Co-O systems~(so-called Co232), 
which had been believed to have the same structure as Bi2212 superconductor.~\cite{Tarascon89} 
But recently, we have revealed~\cite{Yamamoto00} that it is a misfit-layered compound 
isomorphous to [Bi${}_{0.87}$SrO${}_2$]${}_2$[CoO${}_2$]${}_{1.82}$ 
(Pb is undoped) reported by Leligny {\it et al.}~\cite{Leligny99,Leligny00}~(Fig.~\ref{cryst}(a)). 
In this material, Co ions are very stable in the low-spin state as suggested by the 
susceptibility~\cite{Tsukada98} and photoemission spectroscopy measurements.~\cite{Mizokawapre} 
The valence of Co ion is calculated to be +3.33, so that even the Pb-free parent compound should be 
considered as a hole-doped system, and as a consequence, shows metallic behavior.~\cite{Yamamoto00} 

Interestingly, these layered cobalt oxides with large thermoelectric power 
have hexagonal CoO${}_2$ layers in common, 
where Co ions form a two-dimensional triangular lattice as shown in Fig.~\ref{cryst}(b). 
Actually, they show many common features, for example, 
Curie-Weiss-like susceptibility,~\cite{Masset00,Tanaka94,Tsukadapre} 
temperature-dependent Hall coefficient~\cite{Terasaki99,Yamamoto99} and 
negative magnetoresistance.~\cite{Masset00,Tsukada98,Terasaki99} 
Thus, it seems natural that this structure plays an essential role in the 
large thermoelectric power observed in these Co oxides. 
However, the origin of the large thermoelectric power with metallic conduction 
has not yet been so clear, and one of the main reasons for this would be the insufficiency 
of the comprehension of fundamental properties of these systems. 
These materials show both metallic conduction and Curie-Weiss-like susceptibility, 
but even this is not trivial because only $t{}_{2g}$ orbitals seem to be involved with these properties. 
Thus a comparative study of these Co oxides is important, which will elucidate 
the essence of the large thermoelectric power. 

In this paper, we report the measurements of transport~(including magnetotransport) property, 
magnetic property, and specific heat for (Bi,Pb)-Sr-Co-O, 
using polycrystalline and single-crystal samples with different Pb concentrations. 
(Bi,Pb)-Sr-Co-O is an intriguing material because we can control its transport and 
magnetic properties by the content of Pb doping.~\cite{Tsukada98,Tsukadapre,Yamamoto99} 
Heavily Pb-doped Bi-Sr-Co-O shows magnetic transition at low temperatures, 
accompanied by large negative magnetoresistance. 
Thus, we can say that the (Bi,Pb)-Sr-Co-O is the best material to study the coupling between 
charge and spin in the triangular CoO${}_2$ lattice. 

\section{Experimental}
The polycrystalline samples of Bi${}_{2-x}$Pb${}_x$Sr${}_2$Co${}_2$O${}_z$ 
($0 \le x \le 0.4$) were prepared by solid state reactions. 
As reported previously,~\cite{Yamamoto00} we can obtain 
single phase samples by this starting composition. 
The appropriate amounts of high-purity Bi${}_2$O${}_3$, Pb${}_3$O${}_4$, SrCO${}_3$, and 
Co${}_3$O${}_4$ powders were weighed to produce a mixture of the cationic composition 
(Bi,Pb)${}_2$Sr${}_2$Co${}_2$ and then preheated at 
700${}^{\circ}$C in air for 12 h. 
The resultant mixture was ground again and reacted at 800${}^{\circ}$C in flowing 
O${}_2$ for 24 h twice with an intermediate grinding. 
For the resistivity and magnetization measurements, obtained powders were pressed into pellets and 
sintered for another 24 h under the same condition. 

The single crystals were prepared by a floating-zone~(FZ) method 
at a feeding speed of 0.2 mm/h in a flow of O${}_2$. 
The appropriate amounts of high purity Bi${}_2$O${}_3$, Pb${}_3$O${}_4$, 
SrCO${}_3$, and Co${}_3$O${}_4$ powders were weighed to produce a 
mixture of the cationic composition (Bi,Pb)${}_2$Sr${}_2$Co${}_2$. 
The mixture was preheated at 700${}^{\circ}$C in air for 12 h 
and the resultant powder was ground again and sintered at 800${}^{\circ}$C in air 
for 24 h. Obtained powder was pressed into a rod with a typical size of 
5 mm$\phi \times$ 100 mm and sintered at 800${}^{\circ}$C for 50 h in air. 
As the resultant rod is porous, it is necessary to melt it with high feeding 
speed of 100 mm/h before the final growth. 
Obtained single crystals are easy to cleave and platelike. 

Samples were characterized by electron diffraction (ED), and x-ray diffractions (XRD) 
using Cu and Mo $K{\alpha}$ radiations, respectively.
Analysis of the actual composition of the single crystals was made by 
inductively coupled plasma-atomic emission spectroscopy (ICP-AES) technique. 
We have not determined the oxygen content in this study. 
The details of the characterization has been previously reported.~\cite{Yamamoto00} 
In this paper we use four single-crystal samples with different Pb concentrations. 
We summarize the result of the characterization in Table~\ref{samplespec}. 

Single crystals were cut into a rectangular shape with a typical dimension of 
$1.0 \times 0.5 \times 0.05$~mm${}^3$ for the measurement of transport properties. 
The electrical contacts were made with heat-treatment-type silver paint. 
Resistivity measurements were made by a conventional four-terminal technique 
with current parallel ($\rho_{ab}$) and perpendicular ($\rho_c$) 
to CoO${}_2$ planes under the magnetic field up to 9 T. 
Hall-effect measurement was done by four-terminal ac technique. 
To eliminate the contribution of the magnetoresistance due to the asymmetry 
of the electrodes, the magnetic field was changed from -5 to 5 T and the 
Hall voltage was calculated to be $\{V(H)-V(-H)\}/2$, 
where $V$ is the voltage between the Hall probes. 
It is essentially important to make the sample as thin as possible in order to 
obtain the clear data of Hall resistivity, especially at low temperatures where 
carrier localization is observed. 
Magnetization measurement was performed using a commercial superconducting 
quantum interference device~(SQUID) magnetometer 
up to the magnetic field of 7 T. 
AC susceptibility measurement was performed by a mutual inductance method. 
Specific-heat measurement was carried out using a conventional heat-pulse calorimeter 
under the magnetic field up to 13.2~T. 
We also utilized ${}^3$He cryostat for the measurement below 2~K. 

\section{Experimental Results}
\subsection{Basic Properties}
\subsubsection{Transport properties}
\begin{figure}
\epsfxsize=8.5cm
\centerline{\epsfbox{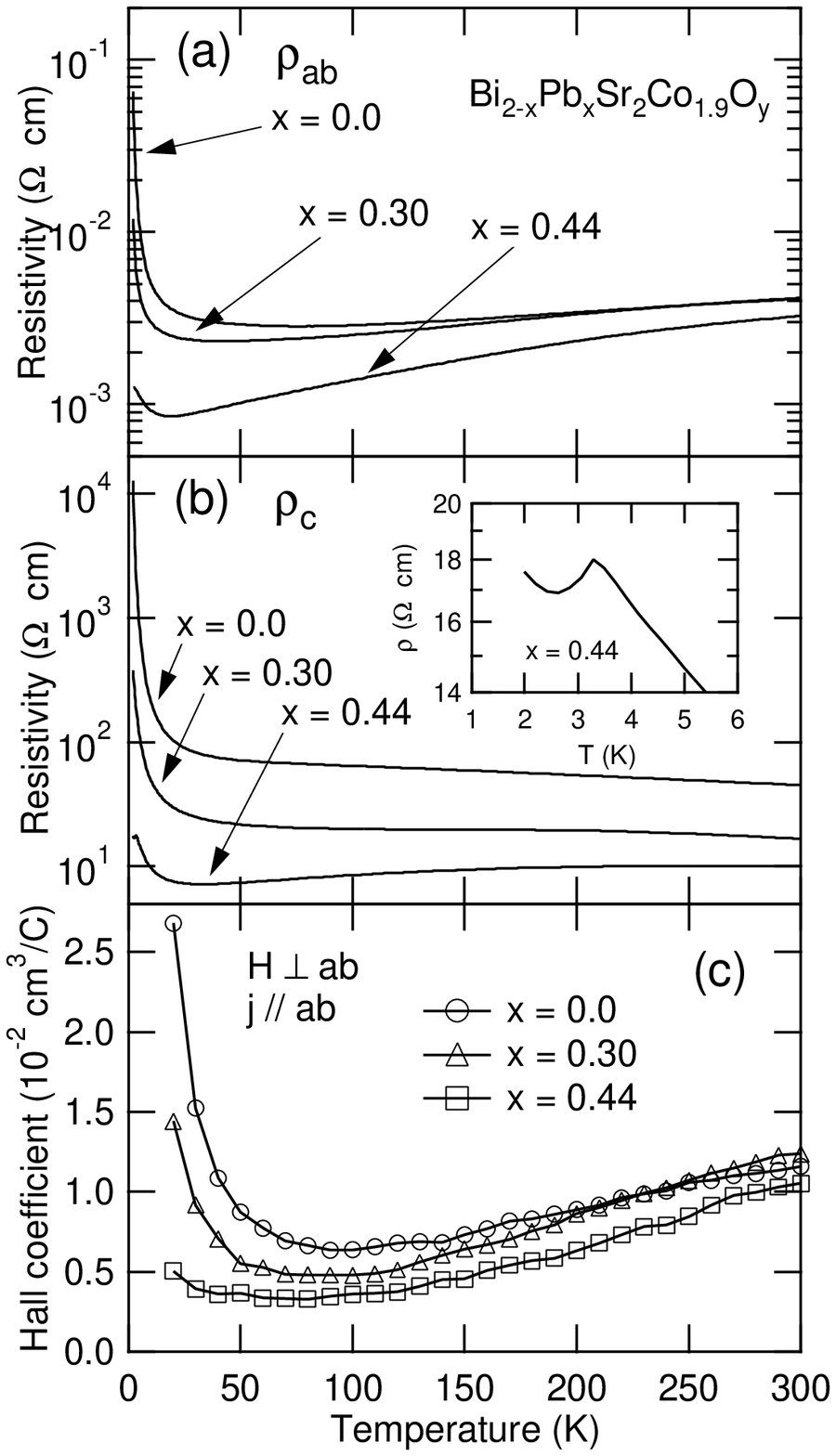}}
\caption{
Temperature dependence of the (a) in-plane resistivity, (b) out-of-plane resistivity and 
(c) Hall coefficient for three single-crystal samples with different Pb concentrations. 
} 
\label{trans}
\end{figure}

\begin{figure}
\epsfxsize=8.5cm
\centerline{\epsfbox{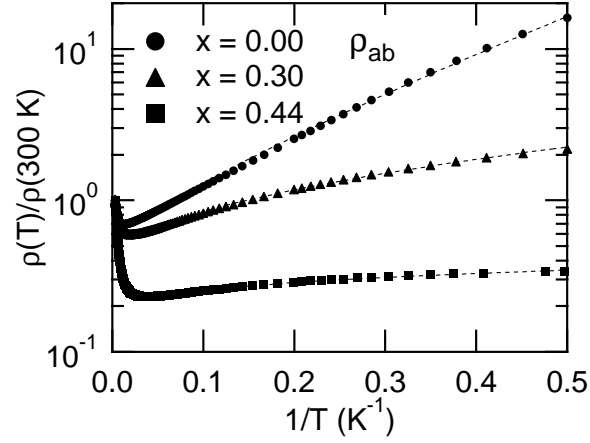}}
\caption{
$\log\rho{}_{ab}$ as a function of $T{}^{-1}$ for three samples. 
The dashed curves are the fits to Eq.~\ref{VRH}. 
} 
\label{Activation}
\end{figure}

Figure~\ref{trans}(a) shows the temperature dependence of the in-plane resistivity ($\rho{}_{ab}$) 
of samples~$x$=0.0, 0.30 and 0.44. Pb-free sample shows metallic behavior around room temperature, 
suggesting the existence of mobile carriers. This result is qualitatively the same as that in 
polycrystalline sample~\cite{Yamamoto00} and reasonable if we consider the valence of Co ions is not +3.0 
in the parent material. However, $\rho{}_{ab}$ of sample $x=0.0$ shows minimum near 80~K 
and diverges with further decreasing temperature. 
The $\rho{}_{ab}$ value at room temperature is around 4 m$\Omega$ cm 
and this is one order of magnitude higher than that of NaCo${}_2$O${}_4$,\cite{Terasaki97} 
but is lower than that of Ca${}_3$Co${}_4$O${}_9$.\cite{Masset00} 
Pb substitution for Bi does not change the $\rho{}_{ab}$ value much at room temperature, but 
it strongly suppresses the divergent behavior at low temperatures and extends the metallic 
region ($d\rho{}_{ab}/dT > 0$), suggesting that carriers are doped by Pb doping. 
To see the conduction mechanism at low temperatures, 
$\log\rho{}_{ab}$ is plotted as a function of $T{}^{-1}$ for the three samples in Fig.~\ref{Activation}. 
As can be seen in the figure, $\rho{}_{ab}$'s of all samples 
show weaker dependence on temperature than that of thermal-activation-type conduction. 
If we assume the variable range hopping~(VRH) conduction, namely, 
\begin{equation}
\log\rho(T) \propto \left(\frac{T_0}{T}\right)^n, \label{VRH}
\end{equation}
$n$ ranges from 0.29 for sample $x$=0.44 to 0.82 for sample $x$=0.0. 
The value of $n$ of sample $x$=0.44~(=0.29) is close to 1/3 that is predicted for 2D VRH conduction, while 
$n$ of the other two samples shows stronger divergence than that of 2D VRH conduction. 

The Pb substitution effect is more drastic in the out-of-plane resistivity ($\rho{}_{c}$) 
as is shown in Fig.~\ref{trans}(b). 
$\rho{}_{c}$ of sample $x$=0.0 increases with decreasing temperature from 300~K and 
diverges as the temperature approaches 0 K. 
The $\rho{}_{c}$ value is about 50 $\Omega$ cm at room temperature and $\rho{}_{c}/\rho{}_{ab}$ 
is $\sim 10{}^4$, suggesting the highly anisotropic electronic structure. 
Pb substitution clearly reduces the overall magnitude 
of $\rho{}_{c}$. It even changes the sign of $d\rho{}_{c}/dT$ 
when we increase the Pb concentration. 
$\rho{}_{c}$ of sample $x$=0.44 shows a broad maximum near 260~K and shows metallic 
behavior down to 30~K, which is consistent with our previous report.~\cite{Tsukadapre} 
Divergent behavior in $\rho{}_{c}$ at low temperatures is also 
suppressed with increasing Pb concentration. 
In the inset of Fig.~\ref{trans}(b), 
$\rho{}_{c}$ of sample $x$=0.44 at low-temperature part is shown. 
A small cusp is clearly observed at the magnetic transition temperature, which will be 
discussed later, and this suggests the strong coupling between the transport and magnetic properties. 
As we reported previously,~\cite{Tsukadapre} 
such a cusp is more clearly observed in $\rho{}_{c}$ than $\rho{}_{ab}$. 

Figure~\ref{trans}(c) is the temperature dependence of the Hall coefficient~($R{}_H$) of 
the three samples. In this figure, Hall coefficients are plotted only in the temperature range 
where the Hall resistivity is linear in the magnetic field below 5.0~T and 
they are defined as proportional constants of the Hall resistivity with the magnetic field. 
$R{}_H$ is positive and strongly temperature dependent in this temperature range. 
The increase of $R_H$ toward the lowest temperature is 
not simply due to the decrease of carrier density, 
but rather due to the anomalous Hall effect, which will be discussed in Section~\ref{AHE}. 

At higher temperature, overall magnitude of $R_H$ is around 0.01 cm${}^3$/C. 
This is more than one order of magnitude higher than that of NaCo${}_2$O${}_4$.~\cite{Terasaki99} 
In NaCo${}_2$O${}_4$, although the strong temperature dependence of $R_H$ is also reported, 
the simple relation $R_H=1/ne$ gives a crude, but a reasonable estimation of 
the carrier density, which is consistent with the result of reflectivity and specific-heat 
measurements.~\cite{Terasaki99} This suggests that the carrier density $n$ of misfit-layered 
Bi-Sr-Co-O system is one order of magnitude smaller than that of NaCo${}_2$O${}_4$. 
If we use the value of $R_H$ of 0.01 cm${}^3$/C, 
the carrier density $n$ is estimated to be of the order of 10${}^{20}$ cm${}^{-3}$. 
This value gives only less than 0.1 holes per Co site,
which is much smaller than 0.33 expected from the chemical composition of 
[Bi${}_{0.87}$SrO${}_2$]${}_2$[CoO${}_2$]${}_{1.82}$.~\cite{Leligny99,Leligny00} 
Carrier density can be also estimated from the reflectivity measurements, 
which had been reported by several authors.~\cite{Terasaki99,Watanabe91,Terasaki93} 
Watanabe {\it et al.} analyzed the reflectivity of 
Bi-Sr-Co-O at room temperature using Drude model~\cite{Watanabe91} and estimated the carrier density to be 
$(m^*/m) \times 3.6 \times$ 10${}^{20}$ cm${}^{-3}$, where $m^*$ and $m$ are the effective masses of carrier and 
free electron, respectively. This is consistent with the above result, if $m^*$ is of the same order as $m$. 
The magnitude of $m^*$ will be discussed later again in the specific-heat measurement. 

The variation of $R{}_H$ with Pb doping is similar to that of the in-plane resistivity. 
Pb doping slightly reduces the magnitude of $R{}_H$, but the increase in the 
carrier number is much smaller than that expected from the chemical composition. 
Probably, the oxygen content is also changed by Pb doping, which compensates the doped carriers. 

\subsubsection{Magnetic properties} 
\begin{figure}
\epsfxsize=8.5cm
\centerline{\epsfbox{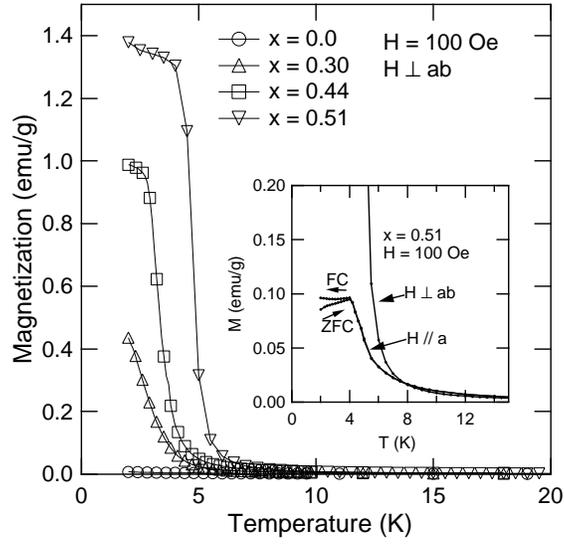}}
\caption{
Temperature dependence of the magnetization under the magnetic field of 100~Oe perpendicular 
to the $ab$ plane. The inset shows the temperature dependence of the magnetization 
under the magnetic field parallel to the $a$ axis. 
} 
\label{M_T}
\end{figure}

\begin{figure}
\epsfxsize=8.5cm
\centerline{\epsfbox{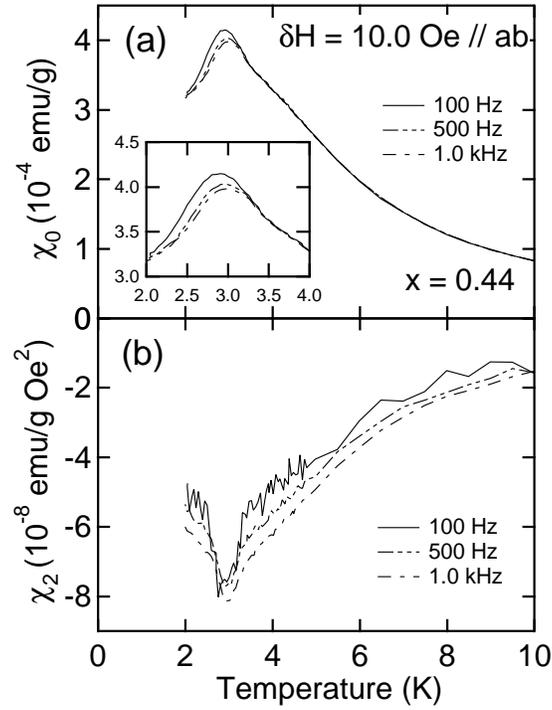}}
\caption[
Temperature dependence of the (a)~susceptibility and (b)~its third harmonics 
of sample $x$=0.44
]
{
Temperature dependence of (a)~ the real part of the fundamental component of the AC 
susceptibility $\chi_0$ and (b)~its third harmonics $\chi_2$ 
of sample $x$=0.44. AC field is applied parallel to the $ab$ plane. 
The inset shows the magnified view near the transition temperature. 
} 
\label{acsus}
\end{figure}

As reported in our previous paper,~\cite{Tsukadapre} 
heavily Pb-doped Bi-Sr-Co-O shows a magnetic transition at very low temperatures. 
In Fig.~\ref{M_T}, we show the temperature dependence of the magnetization under 
the magnetic field of 100 Oe for single-crystal samples. 
The magnetic field is applied perpendicular to the $ab$ plane. 
The magnetization at low temperatures increases with the Pb concentration and 
samples $x$=0.44 and 0.51 show saturated behavior, 
which indicates the magnetic transition in these samples. 
The transition temperature~($T_c$) seems to increase with Pb concentration. 
We will show the variation of the $T_c$ with Pb concentration later. 
On the other hand, the inset shows the temperature dependence of the magnetization 
under the magnetic field of 100 Oe along the $a$ axis in sample $x$=0.51. 
A sharp bend with a small hysteresis is clearly observed around 4~K. 
Such behavior indicates the spin-glass freezing at this temperature. 
To confirm this possibility, we performed AC susceptibility measurement. 
Figure~\ref{acsus} shows the AC susceptibility for sample $x=0.44$ near $T_c$. 
AC magnetic field is applied parallel to the $ab$ plane. 
The cusp at $T_c$ shows clear frequency dependence and the third harmonics show 
anomaly at $T_c$, which are characteristic of the spin-glass system.~\cite{MydoshBook} 

\begin{figure}
\epsfxsize=8.5cm
\centerline{\epsfbox{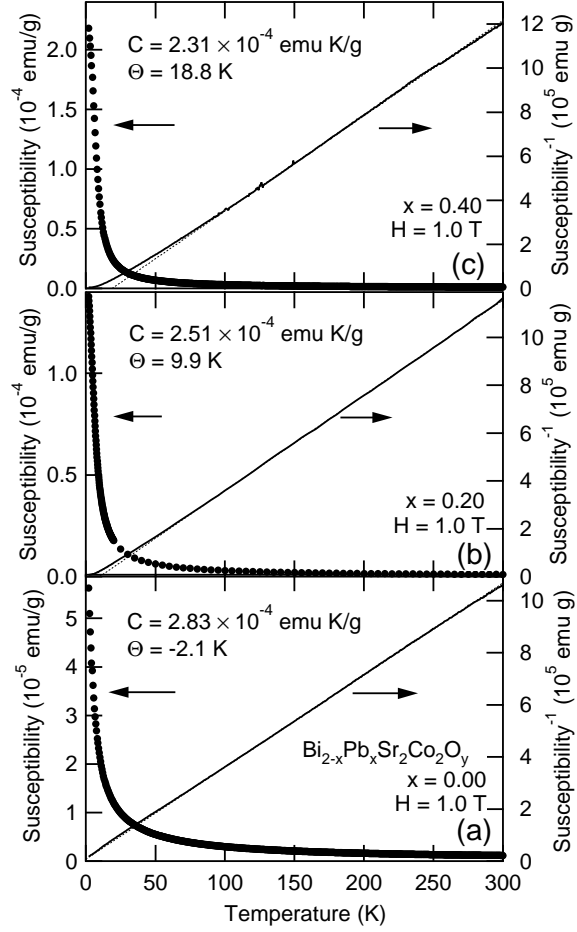}}
\caption{
Temperature dependences of the susceptibility and inverse susceptibility of 
Bi${}_{2-x}$Pb${}_x$Sr${}_2$Co${}_2$O${}_y$ polycrystalline 
samples~($x$=(a)~0.00, (b)~0.20, and (c)~0.40). 
Inverse susceptibility is defined as $(\chi-\chi_0)^{-1}$. 
Dashed lines correspond to the Curie-Weiss behavior. 
} 
\label{poly_sus_T}
\end{figure}

Next, to investigate the spin density and magnetic interaction, 
we measured the temperature dependence of the susceptibility. 
First, we show the data of polycrystalline samples in Fig.~\ref{poly_sus_T} 
because the signal of single-crystal samples is small due to their small mass. 
Here, we fitted the data at 100~K $\le T \le 300$~K for all samples, using the Curie-Weiss law, 
\begin{equation}
\chi(T)=\chi_0+C/(T-\Theta), 
\end{equation}
where $\chi_0$, $C$ and $\Theta$ are temperature-independent susceptibility, Curie constant and 
Weiss temperature, respectively. 
In Fig.~\ref{poly_sus_T}, $(\chi(T)-\chi_0)^{-1}$ is also plotted as a function of temperature. 
As is seen in Fig.~\ref{poly_sus_T}(a), the susceptibility of the Pb-free sample 
follows the Curie-Weiss law well. $\Theta$ is negative, suggesting an antiferromagnetic interaction. 
Let us estimate an effective number of Bohr magneton $p{}_{eff}$ from $C$. 
Since we have not determined the oxygen content of these samples, 
we tentatively assume it to be 8.0.~\cite{comment1}
Then $p{}_{eff}$ is estimated to be 0.97, which leads to 
the existence of low-spin Co${}^{4+}$~($S$=1/2) of 32{\%} of total number of Co ions. 
This concentration is much larger than the value estimated from the Hall coefficient. 

\begin{figure}
\epsfxsize=8.5cm
\centerline{\epsfbox{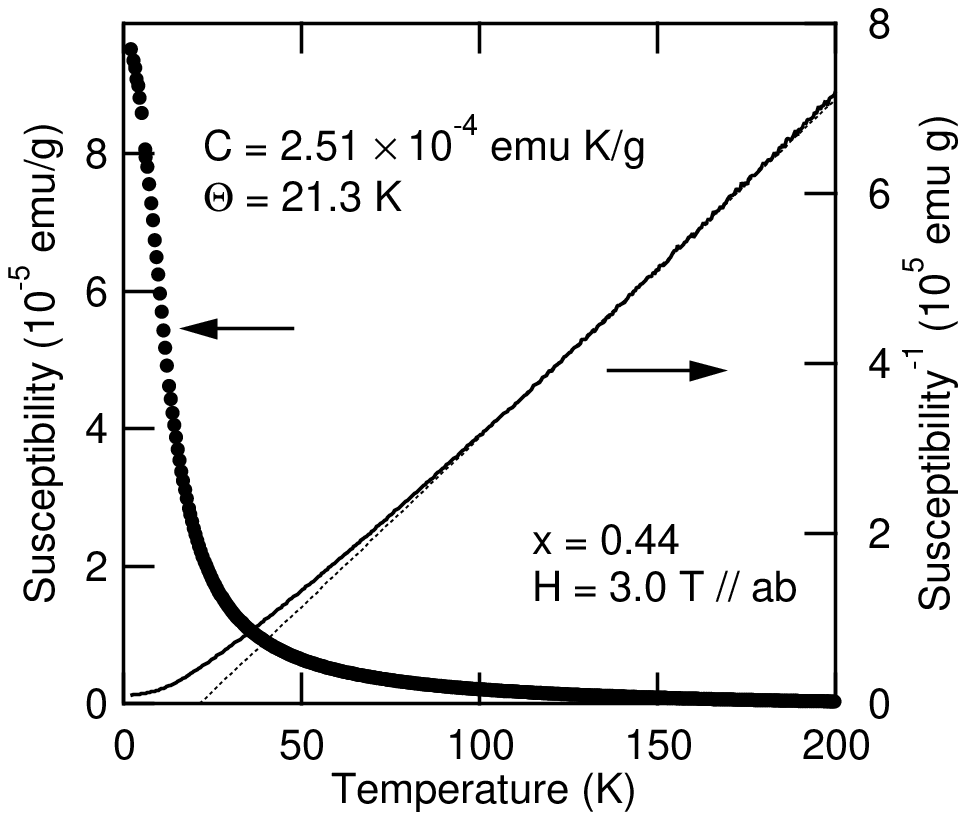}}
\caption{
Temperature dependence of the susceptibility and inverse susceptibility of single-crystal 
sample $x$=0.44 under the magnetic field of 3.0~T parallel to the $ab$ plane. 
Inverse susceptibility is defined as $(\chi-\chi_0)^{-1}$. 
Dashed line corresponds to the Curie-Weiss behavior. 
} 
\label{FZ5_susT}
\end{figure}

When we increase the Pb concentration, the susceptibility clearly deviates from 
the Curie-Weiss law as seen in Figs.~\ref{poly_sus_T}(b) and (c). 
$\Theta$ becomes positive and its magnitude increases with Pb concentration. 
The same behavior is observed in the single-crystal sample. 
Figure~\ref{FZ5_susT} shows the temperature dependence of 
the susceptibility of single-crystal sample $x$=0.44 under the magnetic field of 3.0~T. 
Though we could not measure above 200~K due to the small mass of the sample, 
obtained $\Theta$ is positive and the susceptibility deviates from the Curie-Weiss law 
approximately at 100~K, which is much higher than the actual $T_c$~(3.2~K). 
This behavior is very similar to that widely observed in spin-glass systems 
with short-range magnetic correlation far above the freezing temperature.~\cite{Morgownik81,Rao83} 
Actually, also in the present system, the trace of the short-range magnetic correlation far above $T_c$ 
is observed in magnetotransport and specific heat measurements as will be mentioned in Sections~\ref{Magres} 
and \ref{HeatCap}. 

\subsection{Magnetotransport phenomena}
\subsubsection{Magnetoresistance \label{Magres}}
\begin{figure}
\epsfxsize=8.5cm
\centerline{\epsfbox{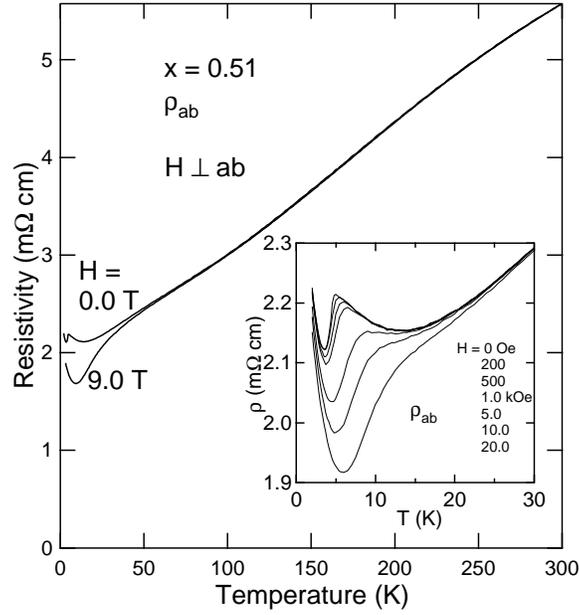}}
\caption{
Temperature dependence of the in-plane resistivity 
of the sample $x$=0.51 under the magnetic fields of 0.0 and 9.0~T. 
The inset shows the magnetoresistance under 9.0~T as a function of temperature. 
} 
\label{FZF_rho_T}
\end{figure}

As was reported in our previous paper, 
(Bi,Pb)-Sr-Co-O shows negative magnetoresistance~(MR).~\cite{Tsukadapre} 
Negative MR is a common feature among NaCo${}_2$O${}_4$,~\cite{Terasaki99} 
Ca${}_3$Co${}_4$O${}_9$,~\cite{Masset00} and (Bi,Pb)-Sr-Co-O, 
but its relation to the magnetism of the system is most clearly observed in (Bi,Pb)-Sr-Co-O. 
Figure~\ref{FZF_rho_T} shows the temperature dependence of the in-plane resistivity~($\rho_{ab}$) 
of the sample $x$=0.51 under the magnetic fields of 0.0 and 9.0~T. 
The magnetic field is perpendicular to the $ab$ plane. 
At low temperatures, negative MR can be clearly observed. 
In the inset, the magnification of the low-temperature part under various magnetic fields is shown. 
At zero field, the resistivity shows a cusp at 5.0~K, 
which corresponds to the magnetic transition. 
With increasing magnetic field, the cusp shifts toward higher temperature. 
Such behavior is commonly observed in the perovskite manganites~\cite{Tokura94} 
and cobalt oxides.~\cite{Yamaguchi95} We will discuss its microscopic mechanism in Section~IV. 

\begin{figure}
\epsfxsize=8.5cm
\centerline{\epsfbox{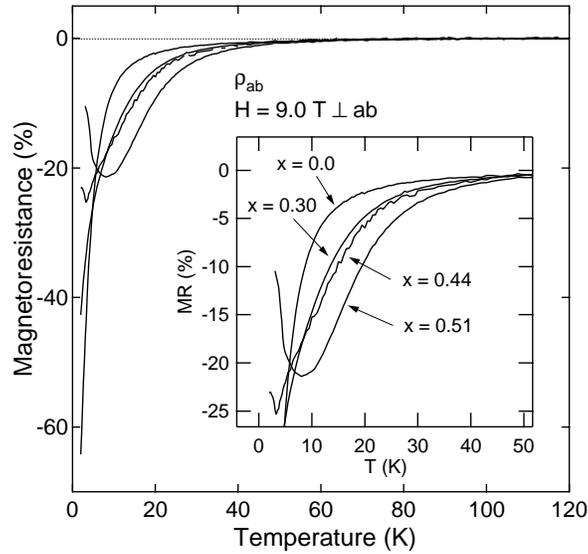}}
\caption{
Temperature dependence of the magnetoresistance under the magnetic field of 9.0~T 
for samples with different Pb concentrations. 
The inset shows the magnification of the low-temperature part. 
} 
\label{MR}
\end{figure}

Next, we show the temperature dependence of the in-plane MR under the magnetic field of 9.0~T 
for four different samples in Fig.~\ref{MR}. 
The magnetic field is perpendicular to the $ab$ plane. 
In all the samples, including sample $x$=0.0 which does not show the magnetic transition above 2~K, 
large negative MR is observed. The temperature below which the negative MR is observable 
increases with Pb concentration. In sample $x$=0.51, negative MR can be observed at much higher 
temperature than its $T_c$=4.5~K. This behavior is consistent with the result of the susceptibility 
measurement. The deviation from the Curie-Weiss law shown in Fig.~\ref{poly_sus_T} suggests 
the existence of the short-range magnetic correlation far above $T_c$ and the deviation becomes 
more remarkable with the increase of Pb concentration. 
The ferromagnetic fluctuation, which is expected from the positive $\Theta$ at high temperature, 
is suppressed by the magnetic field which gives rise to the observed negative MR. 

\subsubsection{Anomalous Hall Effect \label{AHE}}
\begin{figure}
\epsfxsize=17cm
\centerline{\epsfbox{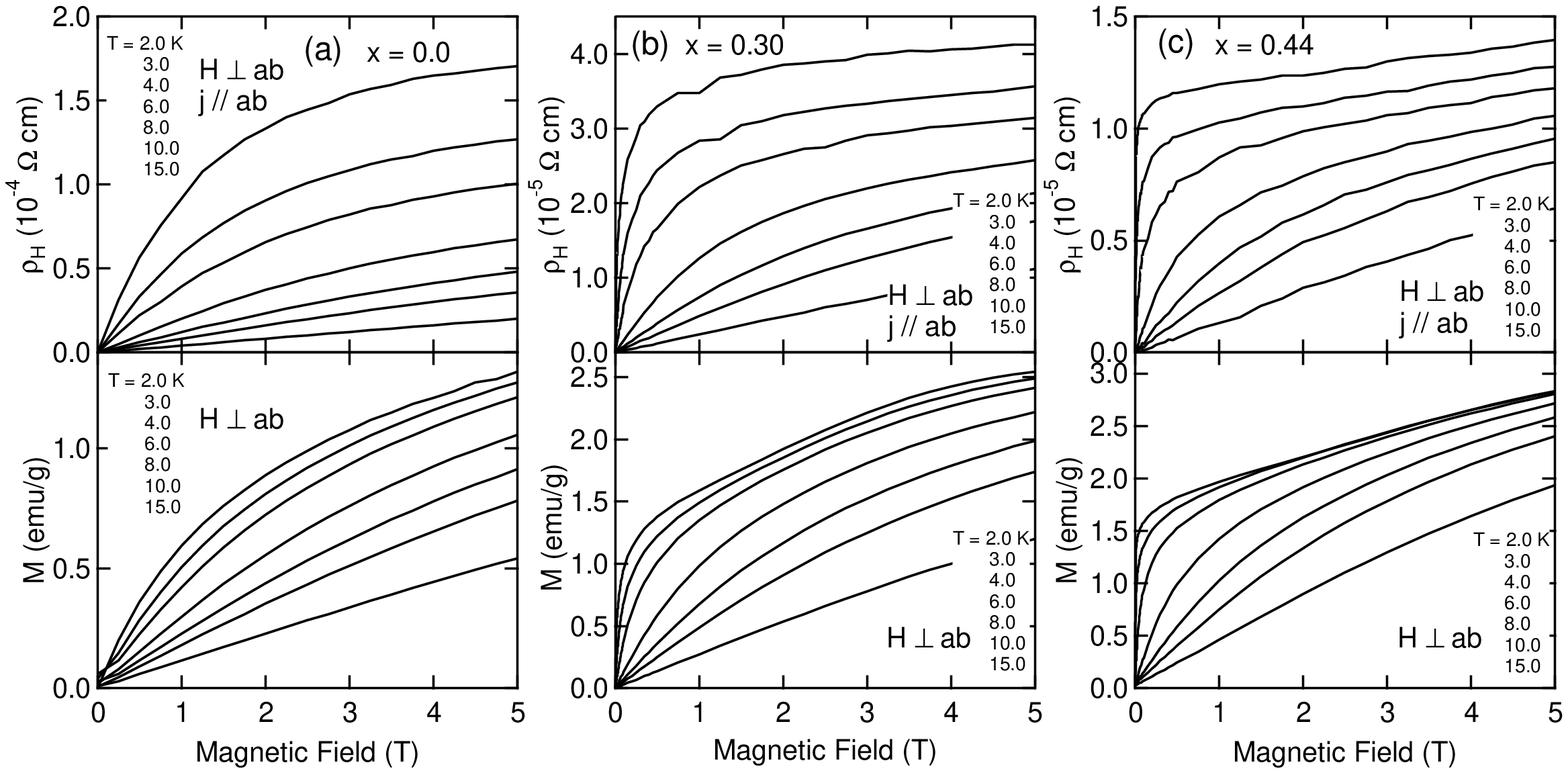}}
\caption{
Magnetic field dependence of the Hall resistivity~$\rho_H$ (upper panel) 
and the magnetization~$M$ (lower panel) 
of samples (a) $x=0.0$, (b) $x=0.30$ and (c) $x=0.44$. 
The data and measuring temperatures correspond in the same order. 
} 
\label{MH}
\end{figure}

As mentioned before, Hall resistivity $\rho_H$ well above the $T_c$ is linear in magnetic field. 
However, if the temperature is decreased approximately 
below 20~K, $\rho_H$'s of all of the samples show nonlinear behavior in the magnetic field. 
Figure~\ref{MH} shows the magnetic field dependence of $\rho_H$ 
(upper panel) and the magnetization $M$ (lower panel) of the three samples below 15.0~K. 
The current and magnetic field are parallel and perpendicular to the $ab$ plane, respectively. 
The measurement of $\rho_H$ in an insulating sample is accompanied by the 
experimental difficulties such as sample heating, large error in Hall voltage 
due to large $|d\rho_{xx}/dT|$, and so on. 
We have confirmed that these effects do not cause essential problems, 
though the magnitude of $\rho_H$ of the most insulating sample $x$=0.0 at 2.0~K 
has a relatively large error less than 10{\%}. 

With decreasing temperature, $\rho_H$ significantly deviates from the 
linear dependence on the magnetic field and shows a profile similar to that of the 
field dependence of $M$ of each sample. This behavior clearly indicates that 
anomalous Hall effect~(AHE) dominates $\rho_H$ in this temperature region. 
Moreover, overall magnitude of the Hall resistivity is very large. 
In particular, $\rho_H$ of sample $x=0.0$ exceeds 100 $\mu\Omega$~cm at low temperatures 
and this is comparable to that of the granular magnetic-nonmagnetic metal alloys 
with much higher resistivity~\cite{Pakhomov96}~(so-called giant Hall effect). 
On the other hand, we can observe a characteristic feature in the $M-H$ curve of sample $x=0.44$, 
that is, the abrupt rise of $M$ with small magnetic field followed by the gradual increase. 
Because of this feature, we cannot simply attribute the linear increase in ${\rho}_H$ 
at high field to $R{}_{0}$, which is commonly used for ordinary ferromagnet.~\cite{HurdBook} 

\begin{figure}
\epsfxsize=8.5cm
\centerline{\epsfbox{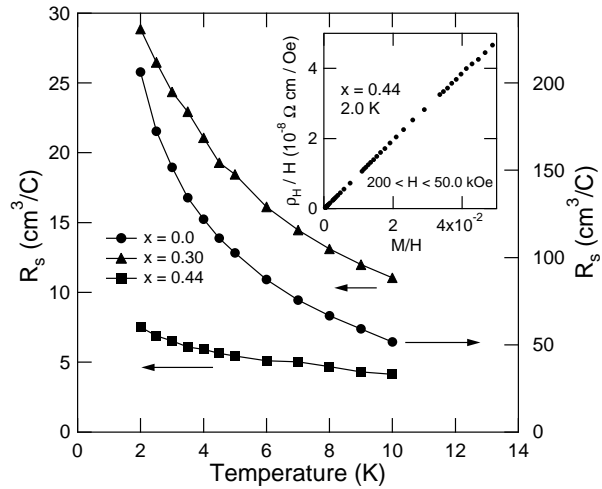}}
\caption{
Temperature dependence of the anomalous Hall coefficient for three samples 
obtained by the fitting~(see text). The inset shows 
the plot of $\rho_H/H$ as a function of $M/H$. 
} 
\label{Rs}
\end{figure}

To estimate the temperature dependence of the ordinary and anomalous Hall coefficients at this 
temperature range, we used the following relation in a magnetic material, which 
has been used conventionally: 
\begin{equation}
{\rho}_{H}=R{}_{0}H+4{\pi}[R{}_{0}(1-N)+R{}_{s}]M, \label{hall}
\end{equation}
where $H$, $R{}_{0}$, $R{}_{s}$ and $N$ are magnetic field, ordinary Hall coefficient, anomalous 
Hall coefficient and demagnetization factor~($N\simeq0.9$ in the present case), respectively. 
$R{}_{0}$ and $R{}_{s}$ are obtained from the plot of ${\rho}_{H}/H$ vs $M/H$ at 1.0 $\le H \le$ 5.0~T. 
The inset in Fig.~\ref{Rs} shows the plot of ${\rho}_{H}/H$ vs $M/H$ for sample $x$=0.44 at 2.0~K. 
${\rho}_{H}/H$ is almost linear in $M/H$ at the wide range of the magnetic field, suggesting 
that Eq.~(\ref{hall}) describes the behavior of ${\rho}_{H}$ very well. 
As can be seen in this figure, the magnitude of $R{}_0$ is very small compared to $R{}_s$. 
Actually, it is several orders of magnitude smaller than $R{}_s$ and is 
sensitive to the range of the data used for the fitting. 
Hence, it is difficult to discuss $R{}_0$ quantitatively and here we concentrate on $R{}_{s}$. 
The main panel of Fig.~\ref{Rs} shows the temperature dependence of $R{}_{s}$ obtained by the fitting. 
$R{}_{s}$'s of the three samples keep increasing down to 2.0~K. 
The magnitude of $R{}_{s}$ of sample $x=0.0$ reaches 200 cm${}^3$/C, which is 
again as large as those of granular magnetic films.\cite{Arozon99} 

Recently, Itoh {\it et al.} pointed out that the present system has some similarities 
to the Kondo semiconductor CeNiSn.\cite{Itoh00} 
They attributed the increasing behavior of $R{}_H$ at low temperatures~(see Fig.~\ref{trans}(c)) 
to the decrease in the carrier number due to some kind of pseudogap formation. 
However, as will be shown later, no trace of the gap formation has been found 
by our specific-heat measurement. Considering the temperature dependences and magnitude 
of $R{}_{s}$'s shown in Fig.~\ref{Rs}, we attribute the increasing behavior of $R{}_H$ to 
AHE. When the magnetization is linear in magnetic field, namely, 
$M$ equals to $\chi H$, Eq.~(\ref{hall}) becomes 
\begin{equation}
{\rho}_{H}/H=R{}_{0}+4{\pi}[R{}_{0}(1-N)+R{}_{s}]\chi, \label{hall2}
\end{equation}
and the second term in this equation contributes to the rising behavior in $R{}_H$. 

Interestingly, even $R{}_{s}$ of sample $x=0.44$ with $T_c$= 3.2~K shows 
monotonic increase with decreasing temperature. 
This behavior of $R{}_{s}$ is in sharp contrast to 
that reported in perovskite manganites or cobalt oxides,~\cite{Wagner97,Asamitsu98,Samoilov98} 
where even in the sample of insulating composition $|R{}_{s}|$ begins to fall off with decreasing
temperature near the magnetic transition temperature. 
We will discuss this problem in Section~IV. 

\subsection{Specific Heat \label{HeatCap}}
\begin{figure}
\epsfxsize=8.5cm
\centerline{\epsfbox{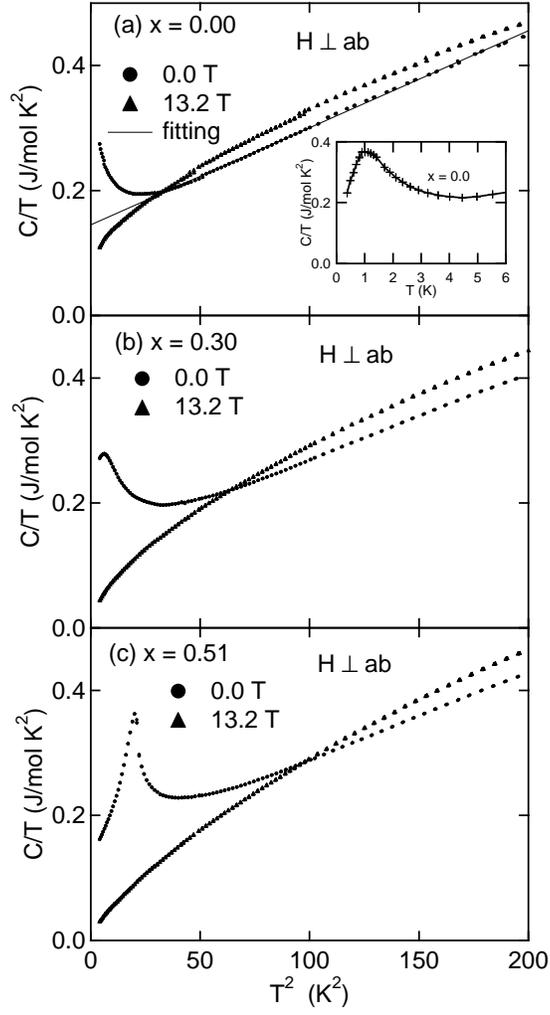}}
\caption{
Specific heat $C$ divided by temperature $T$ as a function of $T^2$ under the magnetic fields 
of 0.0 and 13.2~T for (a) $x$=0.0, (b) $x$=0.30 and (c) $x$=0.51. 
Magnetic field is applied perpendicular to the $ab$ plane. The inset in (a) shows the 
$C/T$ under zero-field condition as a function of temperature for sample $x$=0.0, 
which is obtained by the lower-temperature measurement using ${}^3$He cryostat. 
} 
\label{HC_Pbdep}
\end{figure}

Figure~\ref{HC_Pbdep} shows the specific heat divided by temperature~($C/T$) as a 
function of $T^2$ under the magnetic field of 0.0 and 13.2~T for three single-crystal samples. 
Magnetic field is applied perpendicular to the $ab$ plane. 
$C$ is normalized by the molecular weight determined by the ICP-AES measurement. 
As we do not know the content of oxygen, we assumed it to be 8.0 regardless of the Pb concentration. 
Though it may not be exact, it is only a small correction. 
Even if the content of oxygen changes by 1.0, it gives an error of only about 2{\%} 
for overall magnitude of the specific heat. 

First, we observe a distinct peak at 4.5~K due to the magnetic transition 
in sample $x$= 0.51 under zero-field condition as shown in Fig.~\ref{HC_Pbdep}(c). 
This confirms the existence of the long-range order in this sample. 
For sample $x$=0.30, in which we have not confirmed the magnetic transition above 2.0~K 
by other measurements, we can also observe a peak right above the lowest temperature 
2~K~(Fig.~\ref{HC_Pbdep}(b)). On the other hand, we cannot observe a peak down to 2.0~K 
for sample $x$=0.00 as shown in Fig.~\ref{HC_Pbdep}(a). Thus, we carried out the measurement 
at lower temperature using ${}^3$He cryostat and we observed a peak at 0.90~K as shown in 
the inset of Fig.~\ref{HC_Pbdep}(a). Considering this systematic change of the peak temperature 
in $C/T$ with Pb concentration, it is concluded that even the Pb free samples shows a 
magnetic transition and the transition temperature increases with Pb doping. 

\begin{figure}
\epsfxsize=8.5cm
\centerline{\epsfbox{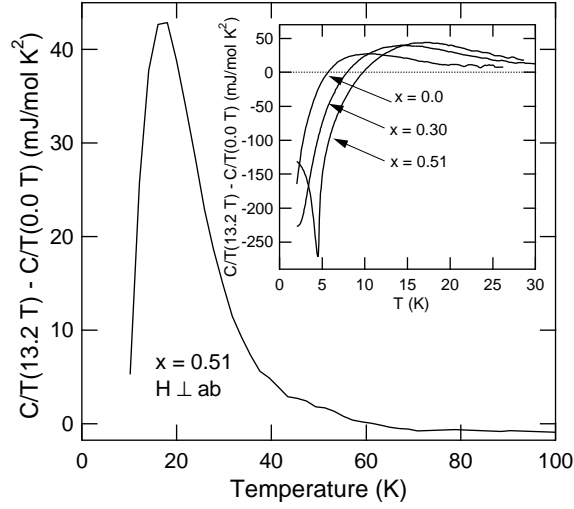}}
\caption{
Temperature dependence of $C/T$(13.2 T) - $C/T$(0.0 T) of sample $x$=0.51. 
The inset shows the temperature dependence of $C/T$(13.2 T) - $C/T$(0.0 T) at low temperatures 
of three samples with different Pb concentrations. 
} 
\label{deltaHC}
\end{figure}

Next, we discuss the magnitude of electronic-specific-heat coefficient $\gamma$. 
As shown in Fig.~\ref{HC_Pbdep}(a), a conventional way to get $\gamma$ by extrapolating 
the high-temperature linear part of $C/T$ to $T$=0 gives a very large a value of 140 mJ/mol K${}^2$. 
However, this is clearly not valid because $C/T$ is strongly suppressed by the application of the 
magnetic field of 13.2~T perpendicular to the $ab$ plane and becomes smaller than the extrapolated line. 
It is well known that a large linear term in $C$ is observed in the disordered spin systems.~\cite{MydoshBook} 
Since the present system also shows spin-glass-like behavior, the observed linear term in $C$ is more naturally 
attributable to this magnetic contribution. 
As shown in Figs.~\ref{HC_Pbdep}(b) and (c), the suppression of $C/T$ at low temperatures 
by the magnetic field becomes more remarkable with Pb doping. 
The suppressed component of $C/T$ seems to shift to higher temperature, 
which, in turn, produces the enhancement of $C/T$ at higher temperature.~\cite{comment3}
As shown in Fig.~\ref{deltaHC}, the enhancement in sample $x$=0.51 persists up to 80.0~K 
under the magnetic field of 13.2~T, which is much higher than the actual $T_c$=4.5~K. 
With decreasing Pb concentration, the enhancement disappears at lower temperature as 
seen in the inset of Fig.~\ref{deltaHC}. 
All these results are naturally understood by the short-range magnetic correlation above $T_c$, 
which is already suggested in the susceptibility and the MR measurements. 
Because of this short-range correlation, the entropy is released gradually 
and the magnetic contribution~($C_{mag}$) to the total specific heat 
exists even at much higher temperature than $T_c$. 
It may be worth referring to the case in CuMn spin-glass system, 
where over 60{\%} of the total entropy possessed by 
Mn spins are released above the freezing temperature.~\cite{Wenger76} 
Since the ferromagnetic fluctuation is strongest in sample $x$=0.51, the variation of 
$C$ with magnetic field is observed to the highest temperature among the samples. 

\begin{figure}
\epsfxsize=8.5cm
\centerline{\epsfbox{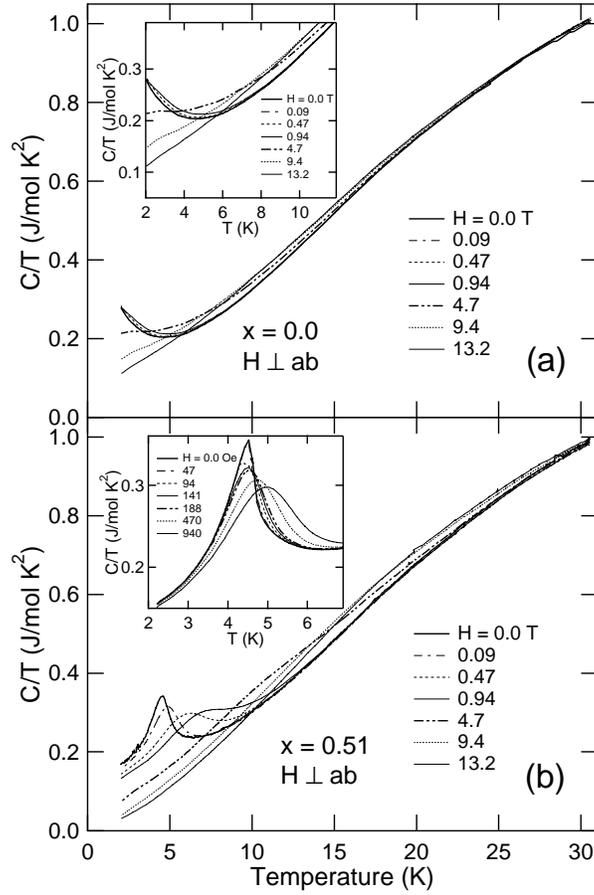}}
\caption{
Temperature dependence of $C/T$ under various magnetic fields up to 13.2~T 
for samples (a) $x$=0.0 and (b) 0.51. 
The insets are the magnified view of the low temperatures part. 
} 
\label{FZEFHC_hdep}
\end{figure}

\begin{figure}
\epsfxsize=8.5cm
\centerline{\epsfbox{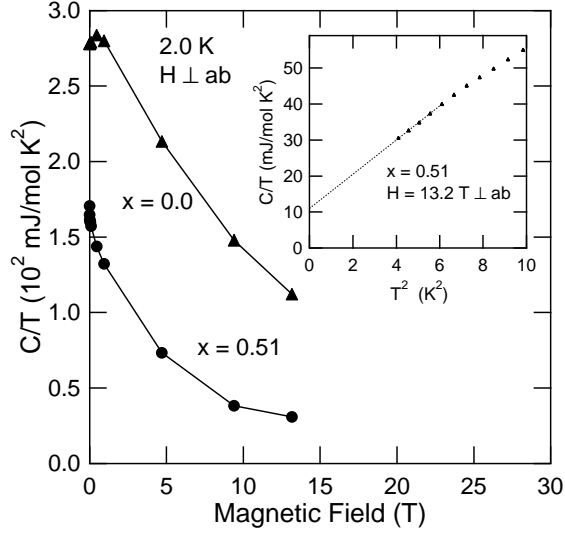}}
\caption{
Magnetic field dependence of $C/T$ at 2.0~K of samples $x$= 0.0 and 0.51. 
The inset shows the $C/T$ of sample $x$=0.51 plotted as a function of $T^2$ under the 
magnetic field of 13.2~T. The dashed line shows the extrapolation of the data to zero temperature. 
} 
\label{CoverT_h}
\end{figure}

In such a situation, however, it is difficult to extract $C_{mag}$ from the total $C$ by the fitting. 
If it is valid to decompose the total specific heat into separate components of $C_{mag}$, 
$C_{ele}$~(electronic term), and $C_{phonon}$~(phonon term), 
we can say that the contribution of $C_{mag}$ below $T_c$ decreases as 
the temperature decreases and the magnetic field increases. 
To get rough estimate of $\gamma$ based on this picture, 
we investigated the magnetic field dependence of $C$ in detail. 
Figure~\ref{FZEFHC_hdep} shows the temperature dependence of $C/T$ of (a) sample $x$=0.0 and (b) 
sample $x$=0.51 under various magnetic fields below 13.2~T. 
The magnetic field is applied perpendicular to the $ab$ plane. 
In sample $x$=0.0, $C/T$ is hardly changed by the magnetic field below 0.09 T. 
When we increase the magnetic field above 0.47 T, the upturn is gradually suppressed and 
it seems that $C/T$ at low-temperature part gradually shifts to higher temperature 
by the application of the magnetic field. On the other hand, in sample $x$=0.51, 
the peak temperature does not depend on the magnetic field below 141~Oe. 
When we increase the magnetic field above 188~Oe, the peak shifts towards higher temperature. 
This behavior is related to the critical field observed in the $M-H$ curve.\cite{Tsukadapre} 
The peak width gradually increases and it eventually collapses approximately at 4.7~T. 
In sample $x$=0.51, the field dependence of $C/T$ becomes weak under high magnetic fields at low temperatures. 
Actually, as shown in Fig.~\ref{CoverT_h}, which is a plot of $C/T$ at 2.0~K as a function of the magnetic field, 
$C/T$ of sample $x$=0.51 shows saturating behavior with the magnetic field of 13.2~T. 
This means that a large part of $C_{mag}$ of sample $x$=0.51 shifts towards higher temperature 
with the magnetic field of 13.2~T, and the magnitude of $C_{mag}$ becomes smaller at 2.0~K. 
Thus, the extrapolation of the data under 13.2~T to zero temperature 
as shown in the inset of Fig.~\ref{CoverT_h} would give a rough estimate of $\gamma$ value 
of 11.0 mJ/mol K${}^2$. 

Using the carrier density estimated from the Hall coefficient, 
this $\gamma$ gives the effective mass of the present system as 6.9 times larger than that 
of a free electron, which is ordinary for a transition metal oxide. 
On the other hand, if we estimate $\gamma$ of sample $x$=0.0 in the same way, 
it becomes much larger value. However, this is clearly invalid because 
$C/T$ of sample $x$=0.0 at low temperatures seems to decrease further by the 
application of stronger magnetic field as shown in Fig.~\ref{CoverT_h}. 
Since the ferromagnetic fluctuation is weak in sample $x$=0.0 as compared with that in sample $x$=0.51, 
$C$ is less sensitive to the application of the magnetic field. 
We have no idea to determine $\gamma$ of sample $x$=0.0 at present. However, considering that 
the ARPES spectra at the Fermi level are hardly changed by Pb doping,~\cite{Mizokawapre} 
$\gamma$ of sample $x$=0.0 is probably similar to that of $x$=0.51. 
Consequently, the effective mass of misfit-layered (Bi,Pb)-Sr-Co-O is not enhanced so much in contrast 
to the case in NaCo${}_2$O${}_4$.~\cite{Ando99} 
This is consistent with the optical reflectivity measurements.\cite{Terasaki99,Watanabe91,Terasaki93} 

From these results of specific-heat measurement, it is strongly suggested that 
misfit-layered (Bi,Pb)-Sr-Co-O has, apart from their origins, localized spin and itinerant 
carrier with not so heavy mass. Based on this picture, 
we will discuss the electronic structure of this system. 

\section{Discussion \label{disc}} 
First, let us consider the origins of the magnetic moments and conductive carriers. 
As confirmed by the susceptibility measurement, about 30{\%} of the Co ions have localized moment of $S=1/2$. 
On the other hand, Hall-coefficient measurement suggests the existence of much smaller number of holes. 
Observed negative magnetoresistance~(MR) and anomalous Hall effect~(AHE) seem to result from the 
interaction between these localized spins and conductive carriers. 
According to the recent x-ray photoemission and absorption spectroscopy measurements 
in misfit-layered (Bi,Pb)-Sr-Co-O,~\cite{Mizokawapre} holes exist mainly in $a_{1g}$ orbitals, 
which are split from $t_{2g}$ orbitals due to trigonal crystal field. 
They claimed that these holes form small polarons due to the strong electron-phonon coupling, 
which serve as localized moments. 
This picture is consistent with our susceptibility measurement because the number of localized 
spins of $S=1/2$ deduced from the Curie constant is near the total number of holes~(33{\%} 
of the number of Co ions), 
which is expected from the chemical formula reported by Leligny and co-workers.~\cite{Leligny99}
Thus, the localized magnetic moment is attributable to this $a_{1g}$ hole. 

Then the problem is the origin of small number of conducting carriers. 
Let us compare two possibilities for this below. 
The first one is that the band derived from Bi-Sr-O layer contributes to the conductivity. 
In Bi-based high-$T_c$ cuprates, band calculation suggests that Bi $6s$ and O $2p$ mixed 
band has dispersion at Fermi level, which contributes to the conductivity.~\cite{Krakauer88} 
Though the Bi-Sr-O layer is highly insulating in the real system of BSCCO superconductors, 
it is possible that this prediction of band theory is realized in misfit-layered Bi-Sr-Co-O 
due to the difference of the crystal structure between these systems. 
However, if carriers originate from the band derived from Bi-Sr-O layers, 
it seems strange that they are strongly coupled to localized spins of Co ions 
as observed in negative MR and AHE. 
Since (Bi,Pb)-Sr-Co-O is a misfit-layered compound, 
Bi-Sr-O and CoO${}_2$ subcells are very weakly coupled to each other. 
Also, as we reported previously, the transport property changes continuously with Pb doping, 
while the lattice constants of Bi-Sr-O subcell changes discontinuously near 
10{\%} of Pb concentration.~\cite{Yamamoto00} This fact makes the above scenario unfavorable. 

\begin{figure}
\epsfxsize=8.5cm
\centerline{\epsfbox{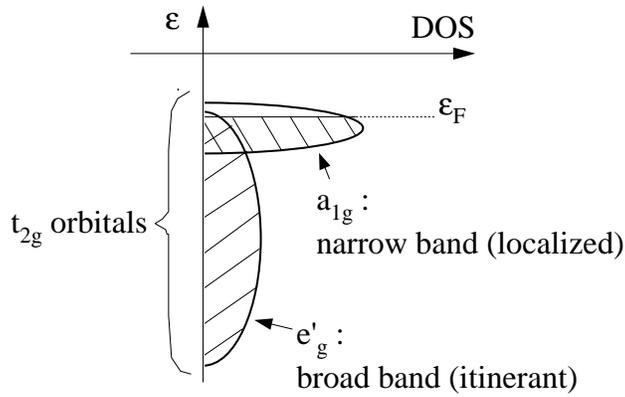}}
\caption{
A model of the band structure of misfit-layered Bi-Sr-Co-O near the Fermi level.
$a_{1g}$ holes are almost localized due to the strong electron-phonon coupling. 
} 
\label{BiSrCo_band}
\end{figure}

The second candidate, which we think more probable, 
is that carriers exist in the $e'_g$ orbitals, which are the other two orbitals 
split from the $t_{2g}$ triplet and spread along the in-plane direction. 
Considering that the number of these carriers is expected to be much smaller, 
and the mass to be lighter than those of $a_{1g}$ holes, 
a band diagram would be like that shown in Fig.~\ref{BiSrCo_band}. 
At the Fermi level, there are two different bands. 
One is, of course, the $a_{1g}$ band, which is responsible for the localized holes 
due to the strong electron-phonon coupling and this is the majority band. 
The other is the $e'_g$ band, which touches the Fermi level and this band is expected to 
provide a small number of holes with relatively small mass. 
Although we do not have a result of the band calculation for this material, 
a similar electronic structure has been discussed in many other systems with trigonally 
distorted $M$O${}_6$ octahedra~($M$: transition metal), such as V${}_2$O${}_3$, 
Ti${}_2$O${}_3$,~\cite{Zeiger75} and LiV${}_2$O${}_4$.~\cite{Anisimov99,Singh99} 
Recently, the band calculation for NaCo${}_2$O${}_4$ has been reported.~\cite{Singh00} 
According to that calculation, the $a_{1g}$ band mainly crosses the Fermi level. 
In addition, the $e'_g$ band hybridized with the $a_{1g}$ band also touches the Fermi level 
to make electron pockets. The band width of the $a_{1g}$ band is smaller than that of the 
$e'_g+a_{1g}$ band. Of course, we cannot apply this result directly to the present system, 
especially because the extent of the splitting between the energy levels of $a_{1g}$ and 
$e'_g$ orbitals are very sensitive to the extent of the distortion of the CoO${}_6$ octahedra.~\cite{Singh00} 
But it seems to be a good starting point to assume this 
type of light and heavy holes at least for the qualitative discussions. 
Based on this model, let us consider other properties in the following.  

First, we consider the magnetic property. 
In our previous paper,~\cite{Tsukadapre} we explained the observed magnetism 
by assuming the canted-antiferromagnetic spin structure. 
However, some of the results, such as the absence of the anisotropy in the in-plane magnetization 
or the positive $\Theta$ in spite of the assumed antiferromagnetic ground state, have remained unclear. 
In this paper, we performed a detailed analysis of temperature and magnetic-field dependence of the 
specific heat, and come to the conclusion that a different picture more naturally explains all the data 
including old ones. It is the coexistence of spin-glass and ferromagnetism. 
Considering the small magnitude of the abrupt increase of magnetization in the 
$M-H$ curve~(0.06 $\mu_B$/Co),~\cite{Tsukadapre} 
a small amount of ferromagnetic cluster with the easy axis of $c$ axis 
is expected to coexist with the spin-glass background. 
The coexistence of the clear cusp and large linear term in the specific-heat measurement 
(see Fig.~\ref{HC_Pbdep}) is consistent with this picture. 
The anisotropic temperature and magnetic field dependences of the magnetization are explained 
as follows: Clear cusps with the hysteresis observed in the temperature dependence of $M_a$ and $M_b$ 
(see Fig.~\ref{M_T} or Fig.~3 in Ref.~\onlinecite{Tsukadapre}) are attributable to the spin-glass freezing, 
while the saturating behavior in $M_c$ is attributable to the ferromagnetic transition. 
Note that $M_a$ and $M_b$ should be isotropic, since the spin glass is isotropic. 
The contribution of the spin glass to $M_c$ is probably smeared due to that of the ferromagnetic component. 
In the M-H curve~(see Fig.~\ref{MH}(c) or Fig.~5 in Ref.~\onlinecite{Tsukadapre}), 
the abrupt increase of $M_c$ with weak magnetic field is due to the ferromagnetic component, 
while the gradual increase at high magnetic field region is 
due to the alignment of glassy spins by the magnetic field. 
The origin of the ferromagnetism is expected to be the double-exchange~(DE) interaction 
between localized spins of $a_{1g}$ holes via conducting $e'_g$ holes. 
This is consistent with the result that the $T_c$ increases with the Pb concentration, namely, 
the enhancement of metallic behavior. 

\begin{figure}
\epsfxsize=8.5cm
\centerline{\epsfbox{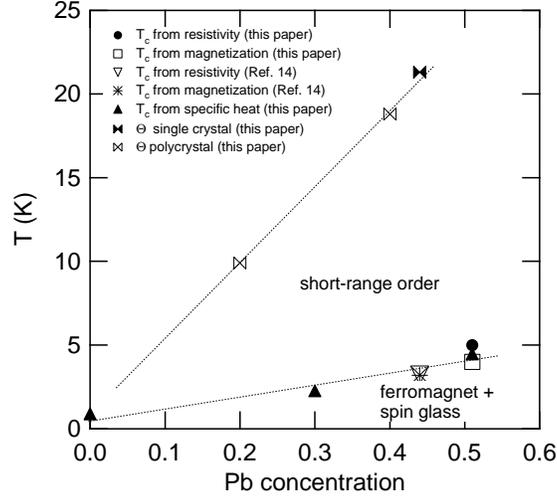}}
\caption{
The magnetic transition temperature $T_c$ and Weiss temperature $\Theta$ 
of misfit-layered (Bi,Pb)-Sr-Co-O system as functions of Pb concentration. 
All $T_c$'s are determined by the measurements for single-crystal samples. 
$\Theta$'s are determined by the 
Curie-Weiss-law fitting as shown in Figs.~\ref{poly_sus_T} and \ref{FZ5_susT} 
for polycrystalline and single-crystal samples, respectively. 
Dashed lines are visual guides. 
} 
\label{T_c}
\end{figure}

There are, however, still some open questions such as the origin of the spin glass. 
Probably, the randomness due to the dilution of the number of spins and/or misfit-layered structure 
is essential, but we need to clarify the origin of the antiferromagnetic interaction. 
Further study is necessary for this problem and we leave here by summarizing the magnetic 
property of misfit-layered (Bi,Pb)-Sr-Co-O system in Fig.~\ref{T_c}, 
where the data in our previous paper are also included. 
It can be seen that the $T_c$ and $\Theta$ increase with the Pb concentration. 
Although it is difficult to say at which temperature the short-range order starts to develop, 
we can say, at least, that it starts to develop at higher temperature when 
the Pb concentration is increased. 

\begin{figure}
\epsfxsize=8.5cm
\centerline{\epsfbox{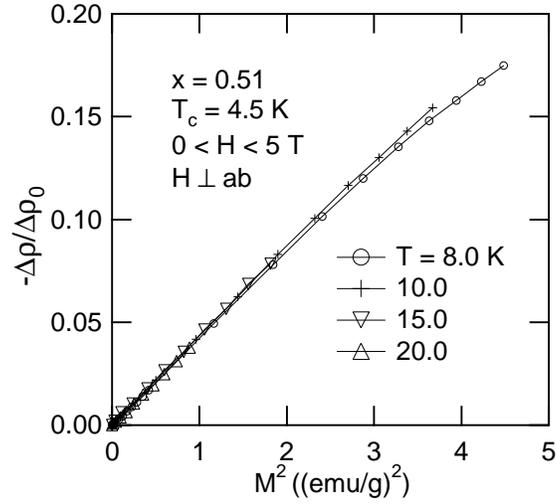}}
\caption{
Magnitude of the in-plane negative magnetoresistance plotted as a function of the squared magnetization at 
various temperatures above $T_c$ for sample $x$=0.51. 
Both magnetoresistance and magnetization are measured under the magnetic field perpendicular to the $ab$ plane. 
} 
\label{MR_scale}
\end{figure}

Next, we consider the magnetotransport properties. 
The origin of the negative MR is expected to be the DE interaction between 
localized spins of $a_{1g}$ holes via conducting $e'_g$ holes. 
It is well known that temperature dependent MR induced by the DE interaction 
can be scaled to a universal curve as a function of the 
magnetization,~\cite{Furukawa94} namely, 
\begin{equation} 
-\frac{{\rho}(H)-{\rho}(0)}{{\rho}(0)} = C M^2, 
\end{equation} 
where $C$ is a constant that measures the effective coupling between the itinerant electrons and 
local spins. This scaling behavior is observed in perovskite manganites~\cite{Tokura94} and 
cobalt oxide.~\cite{Yamaguchi95} 
To check this scaling law, the magnitude of the in-plane negative MR of sample $x$=0.51 is 
plotted as a function of the squared magnetization at various temperatures well above $T_c$ in Fig.~\ref{MR_scale}. 
Both MR and magnetization are measured under the magnetic field perpendicular to the $ab$ plane. 
The scaling holds good at wide range of temperature above $T_c$. 
This result also supports that the DE picture is suitable for the present system. 

The AHE in this system is basically understood by the scattering of $e'_g$ holes by 
localized spins of $a_{1g}$ holes. However, the large magnitude and the temperature dependence of 
$R_s$ are still open questions. The large magnitude is probably related to the magnetic disorder as 
in the granular thin films.~\cite{Pakhomov96} As for the temperature dependence also, 
we point out a possibility that the presence of spin disorder plays an essential role. 
Recently, AHE due to quantal phases in the hopping conduction regime is proposed~\cite{Chun00} 
and discussed in La${}_{1.2}$Sr${}_{1.8}$Mn${}_2$O${}_7$.\cite{Chun00b,Chun00c} 
In such theories, the background spin structure of the system plays an essential role for 
the behavior of the AHE. In the present system, carrier feels the background spin structure 
through Hund coupling. Since spins are disordered and have no periodicity, 
the quantal phase gained by the carrier is expected to have net contribution to the AHE without being canceled. 
This additional contribution may explain the observed increase of $R_s$ down to the lowest temperature 
in sample $x$=0.44. 

Finally, we will briefly comment on the thermoelectric power. 
As is suggested by the reflectivity~\cite{Terasaki99} and specific-heat measurements, 
the mass of carriers in misfit-layered (Bi,Pb)-Sr-Co-O is not enhanced so much, which is in sharp 
contrast to the case of NaCo${}_2$O${}_4$. Thus, at least in misfit-layered (Bi,Pb)-Sr-Co-O, 
mass enhancement does not play an essential role for the large thermoelectric power. 
The imbalance of the spin and orbital degrees of freedom between low-spin 
Co${}^{3+}$ and Co${}^{4+}$ may play an essential role as is theoretically proposed 
by Koshibae and co-workers.~\cite{Koshibae00} 
However, since the carrier density is found to be 
much lower than that of NaCo${}_2$O${}_4$~($\simeq 10^{20}$~cm${}^{-3}$), 
the origin of large thermoelectric power in this system may be different from 
that of NaCo${}_2$O${}_4$. Further study, such as photoemission spectroscopy measurement in 
NaCo${}_2$O${}_4$ and Ca${}_3$Co${}_4$O${}_9$ and comparison with each other is 
highly desired for the elucidation of the physical property of this system. 

\section{Conclusion}
We have investigated the transport property, magnetic property, 
and specific heat in misfit-layered (Bi,Pb)-Sr-Co-O system. 

Overall magnetism and magnetotransport properties of this system 
seem to be governed by the disordered nature of the spin system. 
The spin structure below $T_c$ is found to be understood more naturally 
by the coexistence of spin glass and ferromagnetism rather than the 
canted-antiferromagnetism. 

Susceptibility measurement confirms the existence of nearly 30{\%} of Co${}^{4+}$ among Co ions, 
while the Hall coefficient suggests much smaller number of conductive carriers. 
They are closely coupled to each other, as is manifested in the magnetotransport phenomenon such as 
the negative MR and AHE. 
Specific-heat measurement suggests that the effective mass of the present system is not 
enhanced so much, which is in sharp contrast to the case of NaCo${}_2$O${}_4$ and is consistent 
with the reflectivity measurement reported previously. 
We proposed a two-band model to explain these experimental results consistently. 
The localized spins and itinerant holes are attributed to different bands: 
the former originates from $a_{1g}$ orbitals, while the latter does from $e'_g$ orbitals. 

Recently, search for the superconductivity in the related Co oxides has been reported.~\cite{Loureiro01} 
Though superconductivity has not been observed even under high pressure, 
it seems of great interest to explore other materials with the same type of triangular lattice, 
including non-Co oxides, and investigate their transport, magnetic and thermal properties . 

\section{Acknowledgements}
The authors would like to thank M. Takagi, T. Tsubone, H. Seki, R. Shindo, and S. Tanaka for 
their collaboration at the early stage of this work and S. Ono and Y. Ando for their technical 
advice on transport measurements. 
They would also like to thank 
A. Asamitsu, S. Hebert, T. Kimura, A. Maignan, K. Miyano, T. Mizokawa, 
N. Nagaosa, M. Nohara, B. Raveau, H. Takagi, I. Terasaki, and Y. Tokura for enlightening discussions. 
Some of the measurements were performed at the Cryogenic Center of the University of Tokyo.
This work was partially supported by Grant-in-Aid for COE Research ``SCP-project''. 


\begin{thebibliography}{10}
\bibitem{comment1}
Of course, this assumption is not well justified, because 
the valence of Co is between +3 and +4 and the oxygen content would be between +8 and +9. 
However, this uncertainty hardly affects the estimated value of $p{}_{eff}$ 
because atomic weight of an oxygen is small compared to the total molecular weight.
\bibitem{comment3}
It is worth while to refer to the possibility of itinerant magnetism. 
It is well known that $\gamma$ is enhanced by spin fluctuation 
and $C$ is suppressed by the application of the magnetic field as is observed, e.g., 
in an itinerant ferromagnet Sc${}_3$In.\cite{Takeuchi79} 
Large $C/T$ at zero field and its suppression by magnetic field 
of (Bi,Pb)-Sr-Co-O is qualitatively similar to that of Sc${}_3$In. 
If the present system can be understood as an itinerant magnet, 
it may be meaningless to separate electronic and magnetic components of specific heat. 
However, if this is the case, it seems unnatural that the magnetic transition occurs 
in the localization regime as can be seen in Fig.~\ref{trans}. 
Furthermore, ARPES measurement shows that the DOS at the Fermi level is strongly depleted 
in this system.~\cite{Mizokawapre} 
It is also doubtful that the spin-glass-like behavior observed in the present system 
is explained in terms of the itinerant magnetism. 
Thus, we adopt localized spin picture here to discuss the data of specific heat.
\end{thebibliography}


\begin{table}
\begin{center}
\caption{
Chemical compositions and lattice constants of single-crystal samples.
}
\label{samplespec}
\begin{tabular}{ccccccc}
   Sample & Chemical composition & \multispan{5}{\hfil Lattice constants \hfil} \\
          & Bi : Pb : Sr : Co & $a$~~$b{}_{RS}$~~$b{}_{H}$~~$c$ [\AA],~~$\beta$ [${}^{\circ}$] \\
  \hline
 $x$=0.0 & 2.04 : 0.00 : 2.00 : 1.87 & 4.94~~5.39~~2.8~~14.96,~~93.50 \\
 $x$=0.30 & 1.58 : 0.30 : 2.00 : 1.80 & 4.89~~5.21~~2.8~~15.01,~~93.09 \\
 $x$=0.44${}^*$ & 1.42 : 0.44 : 2.00 : 1.88 & 4.91~~5.23~~-~~15.01,~~92.82 \\
 $x$=0.51 & 1.42 : 0.51 : 2.00 : 1.87 & 4.94~~5.23~~2.8~~15.05,~~92.55 \\
\end{tabular}
\end{center}
* Electron-diffraction observation was not performed in this sample, so that $b{}_{H}$ was not determined. \\
\end{table}

\end{document}